\documentclass[fleqn,twoside]{article}
\usepackage{espcrc2}
\usepackage{graphicx}
\newcommand{\AmS}{{\protect\the\textfont2
  A\kern-.1667em\lower.5ex\hbox{M}\kern-.125emS}}

\title{The Neutron 'Thunder' Accompanying Large Extensive Air Showers}
\author{A.D.Erlykin\address{P.N.Lebedev Physical Institute,   
        Leninsky prosp. 53, Moscow 119991, Russia}
        \thanks{E-mail: erlykin@sci.lebedev.ru}}

\begin{document}

\begin{abstract}
The bulk of neutrons  which appear with long delays in neutron 
monitors nearby the EAS core (~'neutron thunder'~) are produced by high energy 
EAS hadrons hitting the monitors. This conclusion raises an 
important problem of the interaction of EAS with the ground, the stuff of the detectors
 and their environment. Such interaction can give an additional contribution 
to the signal in the EAS detectors at {\em km}-long distances from the large EAS core 
after a few $\mu s$ behind the EAS front.
\vspace{1pt}
\end{abstract}

\maketitle

\section{Introduction}
The dispute on the role of low energy neutrons as the possible origin of delayed 
(~'sub-luminal'~) pulses in the neutron counters and scintillator EAS detectors 
has begun long ago \cite{Tong1,Tong2,Greis,Linsl}. However, the observed delays 
did not exceed a few $\mu s$. The present work has been inspired 
by the observation of the multiple neutrons which followed extensive air showers 
 with delays as long as {\em hundreds} of $\mu s$ \cite{Chub1,Aushe,Anto1,Anto2,Chub2}. 
Such delays have been observed with the Tien-Shan neutron monitor for EAS in the PeV
energy region. Later this finding has been confirmed by other experiments 
\cite{Sten1,Sten2,Gawin,Baygu,Jedrz} and 
the existence of the effect is now beyond any doubts. There is, however, no 
agreement about its origin \cite{Jedrz,Sten3,Chub3}. Briefly the essence of 
the effect is the appearence of the numerous neutrons delayed by hundreds of 
 $\mu s$ after the passage of the main shower disk in the vicinity of the EAS core. 
In the spectacular scenario of 'the thunderstorm model' this phenomenon has been 
compared with the thunder which appears with a delay after the lightning
 during the thunderstorm \cite{Ambro}.  

\section{The PeV energy range}

In the detailed study 
\cite{Anto2} of the effect it has been claimed that the process has a threshold and 
the delayed neutrons appear at PeV primary energies, i.e. in the 
region of the 'knee' observed in the primary energy spectrum. Another observation is 
that these neutrons are concentrated within a few meters around an EAS core and 
accompanied by delayed $\gamma$-quanta. All these
features let the authors assume that this phenomenon is connected with the properties 
of high energy (~PeV~) interactions \cite{Chub3}. 

The other groups \cite{Jedrz,Sten3} attributed the effect to the low 
energy physics, i.e. explained it by neutrons which are produced inside the neutron 
monitor by numerous nuclear scatterings and disintegrations, caused by hadrons in the 
EAS core and which then propagate outside the core region. Some of them appear also as 
albedo neutrons from the nuclear cascade developing in the ground underlying the 
neutron detectors after EAS propagate from the air into the ground. This explanation 
has been based on their own experimental data.  

In \cite{Erlyk} we presented arguments 
based on our simulations at PeV energies which give support to the latter explanation.
Simulations of the EAS neutrons have been also made in \cite{Bouri}. If our interpretation 
of the effect is true, the scenario of 'the neutron thunder' complements our knowledge of the 
EAS development and its interaction with the ground, surface detectors and their 
environment. Like in the case of the transition effect, 
in which some part of invisible gamma-quanta is converted into electrons or electron-positron 
pairs in thick scintillator or water cherenkov detectors this effect demonstrates 
that our records depend on our detectors. Within this scenario another problem appears 
- the production and propagation of neutrons created when the EAS core hits the ground.
 Due to their long propagation length \cite{Agafo} these neutrons can give observable 
effects both at shallow depths underground, in particular at mountain altitudes where the 
EAS cores are more energetic, and also as albedo neutrons - in surface detectors. Due to 
the slow diffusion of thermalized neutrons their contribution to the signal in the 
detectors is relatively high at large delays after the main EAS front. There 
might be plenty of other interesting effects worth of the experimental and theoretical study. 
In this paper we discuss the possible consequence of the effect for the large EAS arrays 
built for the study of cosmic rays (~CR~) at EeV energies. 
  
\section{The EeV energy range}
 
 In the light of the possible 
contribution of neutrons to the signal in water and ice cherenkov 
detectors used in many experiments, viz. Pierre Auger Observatory (~PAO~), MILAGRO, 
NEVOD, Ice-Top and others, we have made simulations of EAS in EeV energy range similar to those 
made at PeV energies \cite{Erlyk}. The interaction model is the same QGSJET-II, but the observation
 level is that of PAO, i.e. 1400 m a.s.l.. The primary proton has 1 EeV energy and the
 vertical incidence angle, the electromagnetic component has been simulated using EGS4 
option with the thinning level of $10^{-5}$. 

The lateral distribution of electromagnetic component (~$e^+ + e^- + \gamma$~), muons 
 (~$\mu^+ + \mu^-$~) and neutrons is shown in Figure 1. I stress that neutrons simulated 
here  are {\em not} secondary neutrons, produced by EAS in the ground, but are those produced 
within EAS in the air. The $\gamma$-quanta are included into the electromagnetic component
 since the water tanks of PAO are thick enough to absorb them and to get the contribution 
to the signal. Both distributions of the particle's number and energy are shown. 
The separation of PAO water tanks is 1.5 km. It is seen that {\em if} neutrons are 
absorbed in the water and their energy is transformed into the visible light they could
 contribute up to 10\% to the signal at such large distances.       
\begin{figure}[htb]
\begin{center}
\includegraphics[width=7.5cm,height=7.5cm,angle=0]{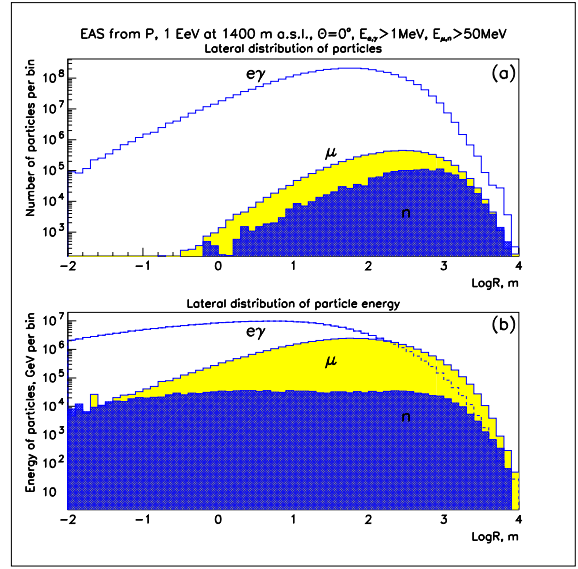}
\caption{\footnotesize Lateral distribution of particle's number (a) and the particle's energy
(b) for 1 EeV primary proton incident vertically at the level of 1400 m a.s.l. It is 
seen that neutrons could contribute up to 10\% to the signal of water tanks at 1km 
distance from the shower axis, if there are relativistic electrons among products of their 
interaction with water.}
\end{center}
\label{fig:neu1}
\end{figure}

It should be remarked that neutrons at large distances from the EAS core have mostly 
the energy below a few GeV and a very wide, nearly isotropic angular distribution. Their 
delays with respect to the shower front spread up to tens of $\mu s$. Interestingly, 
neutrons create two distinct groups with energies above and below $\sim 10^2$GeV. 
Apparently such separation is the consequence of different production mechanisms: 
neutrons above $10^2$GeV are secondaries produced in high energy hadron collisions, 
lower energy neutrons appear mostly in knock-on processes. It is neutrons of the first 
group which carry the bulk of hadron energy together with other hadrons 
in the EAS core. They create $\gamma$- and hadron families in X-ray films and 
ensure the subsequent multiplication process in the neutron monitor. The neutrons of 
the second group diffuse with a non-relativistic speed and with a wide angular distribution to
 the periphery of the shower where they can give delayed 'sub-luminal' pulses in the 
scintillators.

The temporal distribution of the particle's number for electromagmetic, muon and neutron 
component at distances R$<$10m, 100 and 1000 m from the core is shown in Figure 2.
It is seen that after 5$\mu s$ at the core distance of 1km neutrons are the dominant 
component of the shower. Such distances and times are typical for the detectors of PAO. 
 However, neutrons are neutral particles and at these distances they are non-relativistic, 
therefore they cannot emit cherenkov light directly. Only if in the process of their 
moderation and thermalization in water they create relativistic electrons and 
$\gamma$-quanta, they can be detected. In fact such processes do exist, viz. an excitation 
of oxygen nuclei with the subsequent emission of $\gamma$-quanta or 
$n + p \rightarrow d + \gamma$ reaction. The experiment at Tien-Shan confirmed that neutrons 
create such gamma-quanta in the surroundings of the neutron monitor \cite{Anto2}. 
As for water tanks, such a possibility has to be checked experimentally.
In any case the contribution of neutrons into the signal of water cherenkov detectors should 
be estimated and taken into account in the conversion of S(1000), the EAS particle's density 
at 1000m from the core, into the primary EAS energy.     
\begin{figure}[htb]
\begin{center}
\includegraphics[width=7.5cm,height=7.5cm,angle=0]{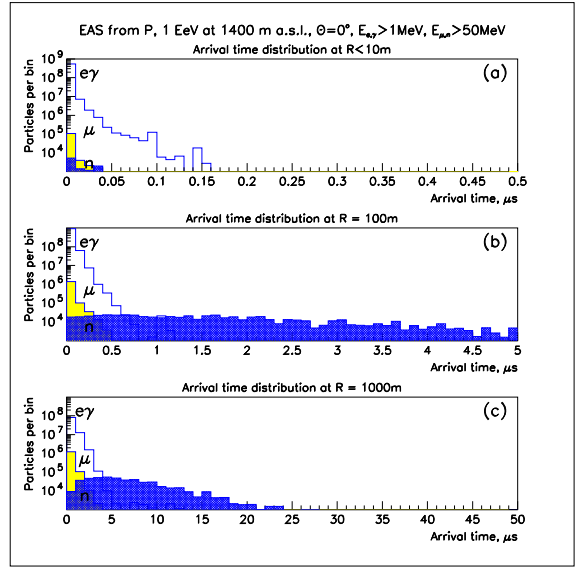}
\caption{\footnotesize Arrival time distribution of electromagnetic, muon and neutron
component of the shower at core distances less than 10m (a), 100m (b) and 1000m (c).
It is seen that at 1000m from the core neutrons dominate among other particles after
5$\mu s$ delay.}
\end{center}
\label{fig:neu2}
\end{figure} 

\section{Discussion and Conclusion}

In 1928 Niels Bohr formulated his 'complementarity principle'. It means that it is 
impossible to observe both the wave and particle aspects of atomic physics 
simulteneously. In other words it implies that only certain information can be gained 
in a particular experiment. Some other information that can be equally important cannot
be gained simultaneously and is lost. This is exactly the situation with the study of
EAS. It is a complex phenomenon and so far different detectors observe and study 
different EAS components: Geiger counters - charged particles, mostly electrons, thick 
scintillators are sensitive also to some part of gamma-quanta, gamma-telescopes - 
cherenkov light, ionization calorimeters - an electromagnetic and hadron component and 
X-rays - highest energy part of these components. Neutrons are neutral particles and so
 far they were not studied separately from all other hadrons. It is a merit of Chubenko
 A.P. and his colleagues who applied the neutron monitor for the detailed study of the 
neutron component of EAS. The neutron monitor is the detector which includes the 
polyethilene as the moderator and reflector - the hydrogen containing material, which increase the 
sensitivity of the device to neutrons. Chubenko A.P. et al. discovered the 'neutron 
thunder' - neutrons delayed up to $ms$ after the passage of the main shower front. 
Alhough according to our interpretation the bulk of the observed neutrons have a 
seconary origin, i.e. they are produced and delayed inside the monitor, the existence 
of the neutrons produced within EAS and accompanying the main shower front is now 
without any doubt. The true 'neutron thunder' associated only with EAS is not so long 
as that observed inside the monitor - the simulations show that it can last up to 
hundred $ns$. However neutrons of EAS can definitely cause the same effects in the
environment, in the ground and in the detectors as they make in the neutron monitor
, like an 'echo effect', which lasts up to hundreds of $\mu s$.
 
First of all it is particularly true for the studies at the mountain level where the EAS core 
as the neutron's producer is more energetic, than at sea level. Secondly a good part of 
the year the ground at the mountain level is covered by such neutron moderator as snow  
(~Tien-Shan, Aragats, Chacaltaya, South Pole~) sometimes a few meters thick. Snow  As for 
the Tien-Shan station there might be an additional factor emphasizing the role of neutrons - 
its ground is a permafrost with a good fraction of ice inside. 

As for the detector sensitivity to neutrons, water and ice tanks are particularly worth of 
attention. First of all water is also a moderator. Secondly, although the neutrons cannot produce 
cherenkov light directly, the study \cite{Anto2} showed that they produce gamma-quanta and 
electrons, which can be eventually detected by water tanks due to their emission of cherenkov light. 
Since water and ice filled detectors are wide spread all over the world and in particular 
used in the PAO, the contribution of neutrons to their signals at large 
distances from the EAS core and at large delays from the trigger time, can be 
substantial. It should be analysed and taken into account if necessary. The same remarks 
could be referred to hydrogen containing plastic scintillators used in many other large
EAS arrays (~Yakutsk, Telescope Array etc.~). 

In any case the phenomenon of 'the neutron thunder' complements our knowledge of the EAS development 
and is certainly worth of the further experimental and theoretical study. 

{\bf Acknowledgments}

The author thanks the INFN, sez. di Napoli and di Catania, personally Professors 
M.Ambrosio and A.Insolia for providing the financial support for this work and their 
hospitality. I also thank Martirosov R., Petrukhin A., Ryazhskaya O.G., Stenkin Yu.V., 
Szabelski J., Ter-Antonian S., Vankov Kh., Watson A. and Yodh G. for useful discussions 
and references.

\end{document}